\documentclass[5p]{elsarticle}
\usepackage{graphicx}
\usepackage{txfonts}
\usepackage{color}
\usepackage{natbib}
\bibpunct{(}{)}{;}{a}{}{,}

\begin{document}

\begin{frontmatter}

\title{N$_2$-associated surface warming on early Mars}



\author[dlr]{P. von Paris\corref{cor}}
\author[zaa]{J.L. Grenfell}
\author[dlr,zaa]{H. Rauer}
\author[dlr]{J.W. Stock}

\cortext[cor]{Corresponding author: email philip.vonparis@dlr.de, tel. +49 (0)30 67055 7939}

\address[dlr]{Institut f\"{u}r Planetenforschung, Deutsches Zentrum
f\"{u}r Luft- und Raumfahrt (DLR), Rutherfordstr. 2, 12489 Berlin, Germany}
\address[zaa]{Zentrum f\"{u}r Astronomie und Astrophysik (ZAA), Technische
Universit\"{a}t Berlin, Hardenbergstr. 36, 10623 Berlin, Germany}

\begin{abstract}

Early Mars may have had a warmer and denser atmosphere allowing for the presence of liquid water on the surface. However, climate model studies have not been able to reproduce these conditions even with a CO$_2$ atmosphere of several bars. Recent 3D simulations of the early Mars climate show that mean surface temperatures only slightly below 273\,K could be reached locally.

We want to investigate the effect of increased partial pressures of N$_2$ on early Mars' surface temperature by including pressure broadening of absorption lines and collision-induced N$_2$-N$_2$ absorption.

A 1D radiative-convective cloud-free atmospheric model was used to calculate temperature profiles and surface conditions. We performed a parameter study varying the N$_2$ partial pressures from 0 to 0.5\,bar at CO$_2$ partial pressures between 0.02\,bar and 3\,bar. These values are consistent with existing estimates of the initial, pre-Noachian reservoir for both species. Solar insolation was set to be consistent with the late Noachian, i.e. around 3.8 billion years ago.

Our 1D global mean simulations clearly show that enhanced N$_2$ content in the Martian atmosphere could have increased surface temperatures. An additional greenhouse warming of up to 13\,K was found at a high N$_2$ partial pressure of 0.5\,bar. Still, even at this N$_2$ partial pressure, global mean surface temperatures remained below 273\,K, i.e. the freezing point of water. However, given the magnitude of the N$_2$-induced surface warming and the results of recent 3D studies which show that local mean surface temperatures are not much lower than 273\,K, our results imply that the presence of atmospheric N$_2$ could have led to almost continously habitable mean surface conditions in some regions. In addition, atmospheric water column amounts increased by up to a factor of 6 in response to the surface warming, indicating that precipitation might also increase upon increasing N$_2$ partial pressure.

\end{abstract}




\end{frontmatter}

\section{Introduction}

There is compelling evidence that the early Martian atmosphere may have been much denser than today. For example, isotope ratio measurements and modelling (e.g., \citealp{mcelroy1976}, \citealp{jak1994}) and inferred impact rates from crater counting \citep{brain1998} suggest intense atmospheric loss during the late Noachian era (see, e.g., review by \citealp{jak2001}). Outgassing through extensive volcanism might have produced a relatively dense CO$_2$ atmosphere of the order of 0.1-1\,bar (e.g., \citealp{phillips2001}). Additional hints for a denser early atmosphere come from the existence of jarosite on the surface \citep{clark2005} or from extrapolated magma ejection trajectories \citep{manga2012}. Together with mineralogical (e.g., \citealp{poulet2005}) and topographical (e.g., \citealp{perron2007} or \citealp{ansan2006}) evidence for liquid water on the surface, these observations indicate a probably warmer, wetter climate on early Mars before 3.8 Gyr ago.

Climate model studies of early Mars have encountered important problems in reproducing these warm and wet conditions with pure CO$_2$-H$_2$O atmospheres (e.g., \citealp{kasting1991}). The potential warming effect of CO$_2$ clouds has been investigated by a number of authors (e.g., \citealp{forget1997}, \citealp{mischna2000}, \citealp{Cola2003}). The conclusion of these studies was that in order to raise the global mean surface temperature above 273\,K, a fractional cloud cover of almost 100\,\% with an associated optical depth of the order of 10 would be needed. These conditions are rather unrealistic, suggesting that the surface of early Mars would remain at temperatures below 260 K. A recent 3D modelling study of early Mars' climate by \citet{wordsworth2013} suggests that even for global mean surface temperatures of the order of 250\,K or lower, locally transient liquid water, albeit in rather small amounts, may be possible. The same study by \citet{wordsworth2013} also showed that, for a 1\,bar CO$_2$ atmosphere, annual mean surface temperatures of the order of $\sim$260\,K are possible in some regions.

An alternative to simple CO$_2$-H$_2$O atmospheres was presented by, e.g., \citet{postawko1986}, \citet{yung1997} or \citet{halevy2007}. They suggested that SO$_2$ concentrations of the order of 10$^{-5}$ would be sufficient to keep the surface temperature above 273 K. Simulations presented by \citet{johnson2008}, using a 3D climate model, confirmed these conclusions and implied that SO$_2$ could act as an efficient greenhouse gas. However, recently, \citet{tian2010} suggested that such high amounts of SO$_2$ in the atmosphere would lead to aerosol formation and subsequent surface cooling.

In the present Martian atmosphere, N$_2$ is the second most abundant constituent. The current N$_2$ partial pressure is about 2$\cdot$10$^{-4}$\,bar. Estimates of the total N$_2$ inventory of Mars range from a few mbars to roughly 0.3\,bar \citep{mckay1989mars}. Assuming a simple scaling to the atmospheric N$_2$ on Venus results in a total N$_2$ inventory of 0.5-0.6\,bar for Mars. Two possibilities have been discussed to explain the fate of the initial inventory of probably several tens or hundreds of mbars. The first involves atmospheric escape. Thermal and non-thermal processes (such as ion pick-up, sputtering etc.) could have eroded around 90\,\% of the initial inventory. By combining all such processes, \citet{jak2001} suggest that more than 99\,\% of volatiles were lost. Statistical events, such as large impacts, could also have significantly eroded the atmosphere. The second possibility to explain the lack of atmospheric N$_2$ on present Mars, apart from atmospheric escape, is incorporation into the crust in the form of, e.g., nitrates. It has been argued that several tens of mbar of N$_2$ could have been fixed as nitrates by, e.g., UV radiation or lightning (e.g., \citealp{manc2003}, \citealp{segnav2005}) and thus be removed from the atmosphere to the (sub-)surface. The detection of such nitrate deposits, if they exist, from orbit is however rather difficult, as has been shown by a Mars-analogue study in the Atacama desert \citep{sutter2007}. Since this study showed that nitrate concentrations increase steeply with depth, in-situ drilling missions are probably needed to assess Martian nitrate deposits. Combining nitrate data with escape signatures from isotopes could then be used to estimate more precisely the actual amount of atmospheric N$_2$ in the Noachian period.

To date, Martian N$_2$ has been mostly of interest in view of a potential biosphere and the possibility of fixing nitrogen in biologically relevant forms (e.g., \citealp{boxe2012}). By contrast, the atmospheric implications of high N$_2$ partial pressures on the climate of early Mars have not been investigated so far. It has been shown for (early) Earth (e.g., \citealp{goldblatt2009faintyoungsun}, \citealp{Li2009}) and the exoplanet candidate GL 581 d \citep{vparis2010gliese} that increasing N$_2$ partial pressures at fixed CO$_2$ partial pressure greatly increases the greenhouse effect, hence surface temperatures. Therefore, in this work, we investigate the influence of increasing N$_2$ partial pressures on the surface temperature of early Mars. We want to assess to what extent an enhanced N$_2$ content could contribute to warmer surface temperatures.

Section \ref{meth} briefly describes
the model and the simulations performed. In Sect. \ref{showresults}, the
results are presented and discussed. Section \ref{conclusions} presents the
conclusions.

\section{Methods}

\label{meth}

\subsection{Atmospheric model}

\label{describemodel}

We used a one-dimensional radiative-convective model as described in \citet{vparis2008,vparis2010gliese}. The model calculates globally,
diurnally averaged atmospheric temperature and water profiles for
cloud-free conditions. In its original form, the model was introduced by
\citet{kasting1984water,kasting1984}. Model updates and improvements to radiative transfer and the treatment of convection are described in, e.g., \citet{kasting1991}, or \citet{mischna2000}. More details on model equations and assumptions are given in \citet{vparis2008,vparis2010gliese}.

Model atmospheres are assumed to be composed of CO$_2$, N$_2$ and H$_2$O only, and divided into 52 model layers. These layers are spaced approximately equidistantly in $\log$\,(pressure), ranging from the surface to the fixed model lid at 6.6$\times$10$^{-5}$\,bar. The model calculates the radiative temperature profile by solving the radiative
transfer equation and imposing radiative equilibrium. The radiative fluxes are calculated on two separate spectral ranges, one for the stellar radiation (heating rates, mostly in the UV, visible and near-IR) and one for the atmospheric and planetary surface emission (cooling rates, mostly thermal IR). If the calculated radiative temperature gradient is steeper than the adiabatic lapse rate, the temperature profile is adjusted to the wet adiabatic lapse rate. This wet adiabatic lapse rate incorporates either CO$_2$ or H$_2$O as condensing species (see, e.g., \citealp{kasting1988}, \citealp{kasting1991} or \citealp{kasting1993} for a detailed description of convection in the model).

The water profile in the model is calculated based on a relative humidity (RH) distribution and the water vapor saturation pressure. The RH profile is an input to the model (see Sect. \ref{runsummar} for details), whereas the water vapor saturation pressure is calculated according to local temperature. Above the cold trap, the water profile is set to an isoprofile taken from the cold trap value.

The pressure broadening of CO$_2$ lines is accounted for semi-empirically by introducing a CO$_2$ foreign continuum (see \citealp{vparis2008,vparis2010gliese}). We follow the approach of \citet{clough1989} in the approximation of line shape and absorption coefficient $k_{\rm{cont}}$ (in cm$^2$):

\begin{equation}\label{ckd}
  k_{\rm{cont}}(\nu,T)=\nu \tanh(\frac{h \nu c}{2k_BT}) \cdot C_{\rm{foreign}}(\nu)\frac{\rho_f}{\rho_0}
\end{equation}
where $\nu$ is frequency, $T$ temperature, $h$, $c$, $k_B$ Planck's constant, speed of light and Boltzmann's constant, respectively, $C_{\rm{foreign}}$ the continuum (values taken from \citealp{schreier2001} and \citealp{schreier2003}), and $\rho_f$, $\rho_0$ are the density of the broadening gas and a reference density, respectively. The use of Eq. \ref{ckd} is consistent with high-resolution radiative transfer codes as developed by, e.g., \citet{clough1981} or \citet{clough1986}. For the frequency- and temperature-dependant terms in Eq. \ref{ckd}, k distributions on a fixed temperature grid are calculated (see \citealp{vparis2008}).

We also introduced N$_2$-N$_2$ collision-induced absorption (CIA) in the IR radiative transfer part of the model. The N$_2$-N$_2$ CIA shows two distinct bands, the fundamental
band near 4.3 $\mu$m and the rotational band in the far infrared
($\lambda$$>$20 $\mu$m). For nitrogen-rich atmospheres, this could become a very important contribution to the overall IR opacity, hence the greenhouse effect (e.g. on Titan, \citealp{lavvas2008}, \citealp{mckay1989titan}). For the fundamental band, fits to
measurements reported by \citet{lafferty1996} were used, whereas for
the rotational band, absorption coefficients have been taken from
molecular simulations by \citet{borysow1986n2n2}. These standard
approximations have been used for example by \citet{buehler2006} to
assess the accuracy of radiative transfer codes compared to measurements
or by \citet{rinsland2010} for retrieving telluric CO$_2$ profiles
based on satellite data.

\subsection{Model scenarios}

\label{runsummar}

All scenarios assumed a solar luminosity $S$ of $0.75$ times the
present value, representing the approximate value 3.8 Gyr ago
\citep{gough1981}.

The model surface albedo was set to $A_S$=0.21, which is close to the surface albedo of present Mars (e.g., \citealp{kieffer1977}) and allows the model to calculate surface temperatures of the order of 217\,K when assuming present-day conditions (solar insolation, surface pressure).

The RH profile in the troposphere was set to unity in the absence of information on its
vertical distribution. This approach of assuming a constant relative
humidity is consistent with other studies (e.g.,
\citealp{mischna2000,kasting1991} assuming RH=1 and
\citealp{Cola2003} assuming RH=0.5). An RH value of unity
provides an upper limit for the surface temperature since the water
vapor content of the atmosphere is most likely being over-estimated.

We varied the partial pressures of CO$_2$ from 0.02-3\,bar and N$_2$ from 0-0.5\,bar. The values chosen for the partial pressures are consistent with reported estimates of the initial Mars volatile inventory (e.g., \citealp{mckay1989mars}) and modeled volcanic outgassing throughout the Noachian (e.g., \citealp{grott2011}).

\section{Results and Discussion}
\label{showresults}

\subsection{Surface temperatures}

\begin{figure}[h]
 \resizebox{\hsize}{!}{   \includegraphics*{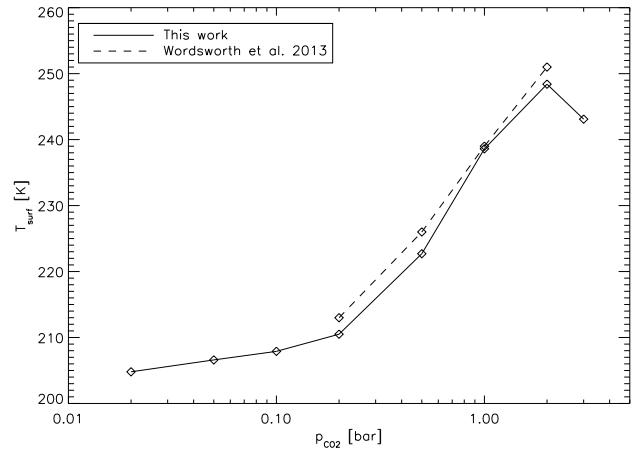}}\\
\caption{Comparison of calculated surface temperatures (at $p_{\rm{N_2}}$=0) with annual mean temperatures from \citet{wordsworth2013} (their Fig. 2, blue points).}
  \label{tcomp}
\end{figure}

Fig. \ref{tcomp} shows the calculated surface temperatures as a function of $p_{\rm{CO_2}}$, for the $p_{\rm{N_2}}$=0 case. For $p_{\rm{CO_2}}$$<$2\,bar, surface temperatures increase with increasing $p_{\rm{CO_2}}$. This is related to the enhanced greenhouse effect. However, for $p_{\rm{CO_2}}$$>$2\,bar, surface temperatures begin to decrease with increasing $p_{\rm{CO_2}}$. This is due to enhanced Rayleigh backscattering of solar radiation, which favors surface cooling (the planetary albedo increases from about 0.2 to 0.45, when $p_{\rm{CO_2}}$ increases from 0.02\,bar to 3\,bar). Hence, the increase in greenhouse warming is compensated by the cooling via increase in Rayleigh scattering, and surface temperatures decrease upon further increasing $p_{\rm{CO_2}}$.

Also plotted in Fig. \ref{tcomp} are annual mean surface temperatures from 3D simulations of early Mars presented by \citet{wordsworth2013}. Surface temperatures agree rather well over the considered range of CO$_2$ partial pressures. Note, however, that we use an approximation of the CO$_2$-CO$_2$ CIA from \citet{kasting1984}, whereas \citet{wordsworth2013} use a more recent parameterization from \citet{wordsworth2010cont}. It was demonstrated by that study that this new parameterization might lead to a cooler surface due to less CO$_2$ opacity. Furthermore, our model is a cloud-free code, whereas \citet{wordsworth2013} include CO$_2$ clouds. In their study, these lead to an increase in surface temperature. Fig. \ref{tcomp} suggests that these two effects approximately cancel.

\begin{figure}[h]
\resizebox{\hsize}{!}{    \includegraphics*{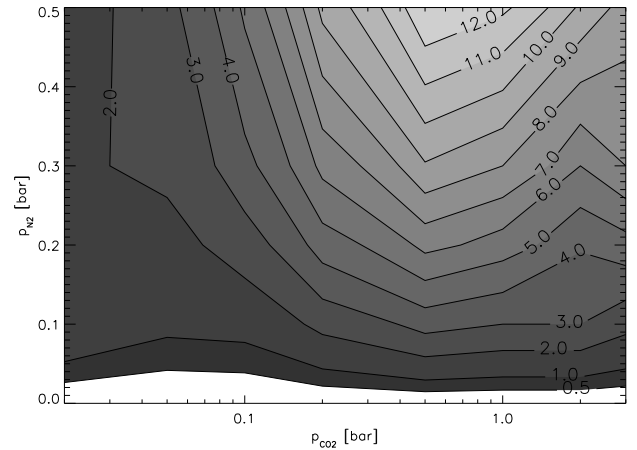}}\\
\caption{Surface warming in K with respect to $p_{\rm{N_2}}$=0\,bar scenarios.}
  \label{ts}
\end{figure}

Fig. \ref{ts} shows the surface warming as a function of $p_{\rm{N_2}}$ and $p_{\rm{CO_2}}$ with respect to the respective $p_{\rm{N_2}}$=0 scenario. It is clearly seen that all scenarios are warmer than the $p_{\rm{N_2}}$=0 case.

The increase of $p_{\rm{N_2}}$ has a significant effect on surface temperatures for CO$_2$ partial pressures of 0.1\,bar and higher, being most pronounced at CO$_2$ pressures of the order of hundreds of mbar. The maximum effect upon increasing $p_{\rm{N_2}}$ to 0.5\,bar is nearly 10\,K (at $p_{\rm{CO_2}}$=2\,bar) and 8.5\,K (at $p_{\rm{CO_2}}$=3\,bar). When increasing $p_{\rm{N_2}}$ to 0.3\,bar (which constitutes the upper limit from estimates in \citealp{mckay1989mars}), the additional greenhouse effect already results in 6 and 7\,K higher surface temperatures, respectively. At 0.5\,bar of CO$_2$, the effect on surface temperature of an increase in N$_2$ is even more substantial, namely 9\,K at $p_{\rm{N_2}}$=0.3\,bar  and 13\,K at $p_{\rm{N_2}}$=0.5\,bar.

An increase of $p_{\rm{N_2}}$ leads to an increase of Rayleigh scattering of incoming solar radiation, hence an increase in planetary albedo and an associated negative radiative forcing. However, since CO$_2$ is a much more efficient scatterer than N$_2$ (e.g., \citealp{vardavas1984}), the radiative forcing due to Rayleigh scattering is rather weak at high $p_{\rm{CO_2}}$. The maximum radiative forcing in the parameter range considered occurs at $p_{\rm{N_2}}$=0.5\,bar and decreases from -7.4\,Wm$^{-2}$ at $p_{\rm{CO_2}}$=0.02\,bar to -1.1\,Wm$^{-2}$ at $p_{\rm{CO_2}}$=3\,bar. Due to higher surface temperatures upon increasing $p_{\rm{N_2}}$, the H$_2$O content of the atmosphere increases (see below), which leads to increased near-IR absorption of solar radiation in the H$_2$O bands. This effect introduces a positive radiative forcing and partly compensates (by about 10-20\,\%) the negative forcing due to Rayleigh scattering.

 At the surface, the increasing $p_{\rm{N_2}}$ leads to a positive radiative forcing for thermal radiation, i.e. an increasing greenhouse effect. This radiative forcing is maximal at $p_{\rm{N_2}}$=0.5\,bar and also decreases with increasing $p_{\rm{CO_2}}$, as does the Rayleigh forcing. At $p_{\rm{CO_2}}$=0.5\,bar, it amounts to 11.6\,Wm$^{-2}$, whereas at $p_{\rm{CO_2}}$=3\,bar it is only 2\,Wm$^{-2}$. This behaviour is due to the fact that at CO$_2$ partial pressures above roughly 1\,bar, the near-surface atmospheric layers are already optically thick at all IR wavelengths, even without the presence of atmospheric N$_2$. Therefore, the surface warming shows a maximum at $p_{\rm{CO_2}}$=0.5-1\,bar. The N$_2$-N$_2$ CIA (see Sect. \ref{describemodel}) contributed less than 1\,K to the increase in surface temperature. Pressure broadening of absorption lines by N$_2$ (e.g., \citealp{goldblatt2009faintyoungsun}, \citealp{Li2009}, \citealp{vparis2010gliese}) constitutes the far more important effect. Further sensitivity tests which varied the amount of H$_2$O and the strength of the CO$_2$-CO$_2$ CIA indicated that the warming effect of N$_2$ is nearly independent of other model radiative transfer details.

Therefore, our calculations suggest that N$_2$ could play an important role for warming early Mars by providing a significant additional greenhouse effect.

\subsection{Implications for liquid surface water and habitability}

Geological and morphological evidence regarding valley networks on Mars implies that during the Noachian, early Mars had a period of extensive fluvial activity (e.g., \citealp{poulet2005}, \citealp{ansan2006}, \citealp{fassett2011}). Climate modeling studies have shown that liquid water could exist temporarily and locally on early Mars even at very low annual mean surface temperatures (e.g., \citealp{wordsworth2013}). The annual amounts of liquid surface water, however, are of the order of 0.1-1\,kg\,m$^{-2}$, i.e. far less than needed to explain the observed fluvial features.

\begin{figure}[h]
\resizebox{\hsize}{!}{    \includegraphics*{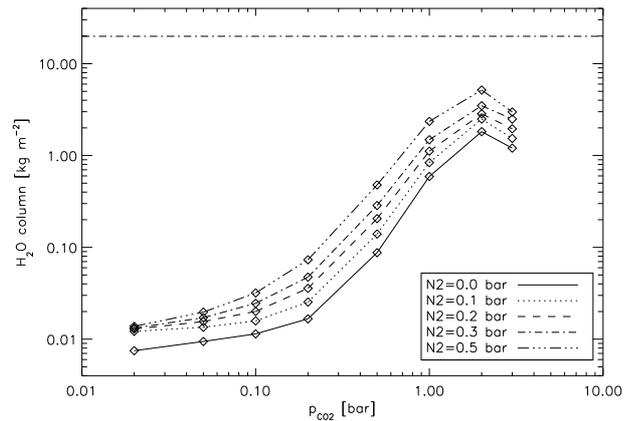}}\\
\caption{Atmospheric H$_2$O column as a function of CO$_2$ and N$_2$ partial pressures. Value of Earth mean column indicated by horizontal line.}
  \label{h2ocol}
\end{figure}

Fig. \ref{h2ocol} shows the atmospheric H$_2$O column, as a function of $p_{\rm{CO_2}}$ for different $p_{\rm{N_2}}$. For reference, we show also the value of the mean atmospheric H$_2$O column on modern Earth as a horizontal line. Note that on modern Earth (1990), $p_{\rm{CO_2}}$=3.5$\cdot$10$^{-4}$\,bar. It is clearly seen that an increase of N$_2$ partial pressure leads to considerably wetter atmospheres, as could be expected from Fig. \ref{ts}. The increase of atmospheric H$_2$O reaches up to a factor of 4-5 (at 1, 2 and 3\,bar of CO$_2$) when increasing N$_2$ to 0.5\,bar, and around a factor of 6 at 0.2\,bar of CO$_2$. Therefore, it seems likely that precipitation and annual surface liquid water could also increase by corresponding amounts. Note, however, that even the considerably increased H$_2$O columns at $p_{\rm{N_2}}$=0.5\,bar only represent about 10$^{-3}$-10$^{-1}$ of the terrestrial atmospheric H$_2$O column. Nevertheless, this increase could help to resolve partly the apparent contradiction between observed early fluvial activity and climate modeling studies.

Also, the calculated increase in global mean surface temperature of up to 13\,K for significant amounts of atmospheric N$_2$ implies that locally, mean annual temperatures could be close to or higher than 273\,K (see, e.g., temperature maps in \citealp{wordsworth2013}). These surface temperatures would allow for liquid water, one of the key requisites for life on Earth. The presence of liquid water is also a fundamental part of virtually all habitability definitions (e.g., \citealp{kasting1993} for terrestrial planets, \citealp{mepag2005} for the Mars exploration context). Therefore, our results imply that early Mars could have been continously habitable at least in some regions. Due to the (chaotic) evolution of both Martian eccentricity and obliquity (e.g., \citealp{laskar2004}), the location of these continously habitable regions might change on timescales of 10$^4$-10$^5$ years (see, e.g., \citealp{wordsworth2013}). Furthermore, it should be noted that with increasing surface pressure, the dependence of surface temperature on topography would most likely also increase (e.g., \citealp{wordsworth2013}), hence complicating the explanation of the mostly high-altitude late Noachian valley networks.

\subsection{Noachian atmospheric N$_2$ content}

Our parameter study aimed at investigating the influence of atmospheric N$_2$ on surface temperatures. Fig. \ref{ts} shows that significant amounts of N$_2$ of the order of 0.3\,bar are needed in the model to warm effectively the surface. Such amounts are towards the upper limit of Martian N$_2$ estimates \citep{mckay1989mars}. Simulations of atmospheric escape suggest that most of the early protoatmosphere was probably lost very shortly after the formation of Mars (e.g., \citealp{tian2009_mars}, \citealp{lammer2013} and references therein). The subsequent re-formation of a secondary atmosphere containing significant amounts of N$_2$ could be possible by, e.g., impact-related delivery (e.g., \citealp{manning2008,manning2009}, \citealp{deniem2012}), since chondrite-type impactors contain relatively high amounts of nitrogen (e.g., \citealp{kerridge1985}, \citealp{hashizume1995}). Volcanic outgassing or nitrate decomposition has also been suggested to contribute to secondary N$_2$ in the atmosphere (e.g., \citealp{owen1977}, \citealp{wallis1989}, \citealp{pepin1994}, \citealp{jak1994}).

Hence, upcoming investigations of the Martian surface and sub-surface (either from orbit or in-situ) should ideally include the search for potentially missing N$_2$. Future atmospheric escape or outgassing studies should also consider N$_2$. In doing so, the present-day Martian N$_2$ reservoir could be constrained much better and it could be assessed whether N$_2$ was present in high enough concentrations in the early Mars atmosphere to influence the Noachian climate.

\section{Conclusions}
\label{conclusions}

We have presented model calculations of the early Mars atmosphere with a 1D cloud-free atmospheric model. We varied N$_2$ partial pressures as well as CO$_2$. To our knowledge, ours is the first study in which the radiative effects of N$_2$ (pressure broadening and N$_2$-N$_2$ collision-induced absorption) were investigated in simulations of the early Martian climate.

Results suggest that increasing N$_2$ partial pressures could lead to significant surface warming, without accounting for CO$_2$ clouds or additional greenhouse gases. Surface temperatures increased by up to 13\,K at high N$_2$ partial pressures of 0.5\,bar. Nevertheless, global mean surface temperatures still remain below 273\,K. Given the results of previous 3D studies of CO$_2$-H$_2$O atmospheres, however, our results suggest that local annual mean surface temperatures above the freezing point of water might be possible. The water column density increased by up to a factor of 6, indicating that precipitation might also significantly increase.

Detailed studies of the outgassing history, impact delivery and atmospheric escape of N$_2$ are needed to assess whether high N$_2$ partial pressures could be consistent with the late Noachian period.

A further investigation of the effect of increasing N$_2$ upon the transient existence of liquid water on the surface and the formation of fluvial features with detailed 3D atmospheric models is warranted in the future.

\section*{Acknowledgements}

This research has been supported by the Helmholtz Association
through the research alliance "Planetary Evolution and Life". We thank A. Borysow for making freely available the Fortran programs to calculate the collision-induced absorption coefficients for N$_2$. We furthermore thank M. Mischna and an anonymous referee for valuable, constructive remarks on the manuscript.

\bibliographystyle{natbib}
\bibliography{literatur_mars}

\end{document}